\documentclass[%
 reprint,
twocolumn,
 amsmath,amssymb,
 aps,
]{revtex4-2}

\usepackage{graphicx}
\usepackage{dcolumn}
\usepackage{bm}

\newcommand{\ket}[1]{\left|#1\right\rangle}
\newcommand{\bra}[1]{\left\langle#1\right|}
\newcommand{\avg}[1]{\left\langle#1\right\rangle}

\newcommand{\pqty}[1]{\left(#1\right)}
\newcommand{\bqty}[1]{\left[#1\right]}

\begin{document}

\preprint{APS/123-QED}

\title{Single-pass stochastic cooling for electron microscopy}%

\author{River Robles}\email{river.robles@fau.de}
\author{Peter Hommelhoff}\email{peter.hommelhoff@lmu.de}

\affiliation{Department of Physics, Friedrich-Alexander-Universit\"at Erlangen-N\"urnberg, 91058 Erlangen, Germany}
\affiliation{Faculty of Physics, Ludwig-Maximilians-Universit\"at, 80799 M\"unchen, Germany}

\date{\today}

\begin{abstract}
Scientific applications of free electrons, such as in electron microscopes and particle accelerators, are limited by the brightness of available electron sources. Accelerators can use phase space cooling to surpass these source limitations, whereas electron microscopes can only rely on particle filtering to increase beam coherence at the cost of reduced brightness. We show that strong electron-light coupling in nanophotonic structures makes it possible to achieve single-pass stochastic cooling inside an electron microscope for a lossless improvement of longitudinal emittance. This approach is laser-free and compatible with normal CW operation of electron microscopes. We derive a fully quantum mechanical model of stochastic cooling valid in the strong coupling regime, finding an analytic asymptotic formula for the optimal cooled energy spread $\sigma_{E_f}\simeq0.88(\hbar\omega)^{2/3}\sigma_{E_0}^{1/3}$, with $\hbar \omega$ the cooling photon energy and $\sigma_{E_0}$ the initial energy spread. Numerical simulations confirm that the energy spread reduction is reflected in an increase in the purity of the electron's quantum state and a reduction of the longitudinal phase space area. Our work paves a path towards even brighter beams for electron microscopy, promising higher image quality, enhanced spectral resolution, and a route to reaching fundamental quantum limits on electron beam phase space area.
\end{abstract}

\maketitle

\textit{Introduction} --- High-brightness electron beams underpin critical scientific instruments from large particle accelerators \cite{musumeci2018advances} to tabletop electron microscopes \cite{spence2013high}. Electron microscopes combine tip emitters with spatial and energetic filters to achieve sub-\AA\ spatial resolution and sub-eV energy widths \cite{batson2002sub,krivanek2014vibrational}. Ultrafast electron microscopes additionally use photoemission to confine electron pulses below one picosecond, typically with about one electron per pulse \cite{zewail2010four,feist2017ultrafast,arbouet2018ultrafast,houdellier2018development,schroder2025laser,chlouba2026controlling}. Tip-based sources are among the closest we have to producing quantum-uncertainty-limited electron beams, yet are still orders of magnitude away from fundamental quantum limits \cite{spence1994electron,kiesel2002observation,kuwahara2014coherence}. Small energy spreads in particular are important both for direct spectroscopy as in electron energy loss spectroscopy and for minimizing chromatic aberrations in imaging. The energy width can be further reduced with monochromators which remove off-energy electrons \cite{tsuno1997design,krivanek2019progress} or with active phase space manipulations such as THz modulation \cite{kealhofer2016all,yannai2023lossless,streshkova2024monochromatization}. Both have disadvantages. The first reduces the usable current, while the second is limited by Liouville's theorem and cannot improve the longitudinal emittance dictated by the electron source.

Phase space cooling techniques avoid both disadvantages by moving entropy from the electron beam into a different system. To our knowledge, no such scheme has been proposed or demonstrated in an electron microscope. In accelerators, on the other hand, techniques like stochastic cooling (SC) \cite{van1972stochastic,mohl1980physics,van1985stochastic} are well-established. In transit-time optical stochastic cooling (OSC) \cite{mikhailichenko1993optical,zolotorev1994transit}, spontaneous radiation emitted in a pickup is used to later remove energy fluctuations from the electron in a kicker. OSC was recently demonstrated at the Integrable Optics Test Accelerator with 100 MeV electron beams \cite{lebedev2021design,jarvis2022experimental}. OSC must generally be performed in a storage ring over many passes due to the large initial energy spread, limited spontaneous undulator radiation emission rates (about $\alpha=1/137$ photons per electron in the fundamental mode \cite{pellegrini2016physics}), and incoherent kicks from other particles in the bunch.

In this Letter, we propose to extend stochastic cooling to electron microscopes. Unlike in accelerators, in microscopes the energy spread to be cooled is at the eV level. This corresponds to no more than tens of (mid-infrared) photons, requiring a fully quantum treatment of the electron and the radiation. Critically, electron microscopes can leverage nanophotonic structures enabling higher electron-light coupling efficiencies than are possible in undulators \cite{garcia2010optical,breuer2013laser,peralta2013demonstration,yang2018maximal,kfir2019entanglements,kfir2020controlling,dahan2021imprinting,feist2022cavity,adiv2023observation,roques2023free,chlouba2023coherent,di2024toward,de2025roadmap,xie2025maximal,zhao2025upper,arend2025electrons}. We show that high coupling efficiency makes single-pass cooling possible, making the scheme naturally compatible with electron microscope columns. This approach is also laser-free, making it compatible with standard CW operation of electron microscopes. We develop a fully quantum mechanical model of the stochastic cooling process in the language of free-electron quantum optics \cite{di2019probing,garcia2021optical,ruimy2025free}, extending earlier quantum treatments of OSC \cite{heifets2001quantum} to the strong-coupling regime. We derive optimal conditions for cooling yielding a minimal energy spread $\sigma_{E_f}\simeq0.88(\hbar\omega)^{2/3}\sigma_{E_0}^{1/3}$, with $\hbar\omega$ the cooling photon energy and $\sigma_{E_0}$ the initial energy spread. With numerical tracking of the electron density matrix, we show that this reduced energy spread corresponds to true phase space cooling, reflected as an increase in the purity of the electron's quantum state and a reduction of its longitudinal emittance. This paves a path towards bringing the already impressive brightness of electron microscopes closer to its fundamental quantum limits, which is of highest practical and fundamental interests.

\begin{figure}[h!]
    \centering
    \includegraphics[width=\linewidth]{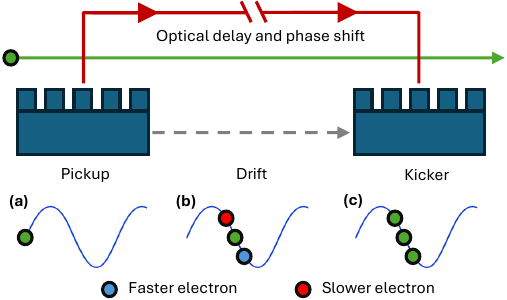}
    \caption{Concept for single-pass stochastic cooling. In a pickup, the electron emits spontaneous radiation whose phase encodes the electron arrival time. After a dispersive drift, the electron meets the radiation in the kicker, where fast (slow) electrons end up in a decelerating (accelerating) phase, cooling the total energy spread. (a), (b), and (c) show the electron-light overlap after the pickup, drift, and kicker, respectively.}
    \label{fig:concept}
\end{figure}

\textit{Theory} --- OSC consists of three stages (Figure~\ref{fig:concept}): (1) the ``pickup" where the electron spontaneously emits radiation, (2) the drift during which the electron dephases from the light depending on its energy, and (3) the ``kicker" in which the light emitted in the pickup modulates the electron energy. The pickup radiation encodes the arrival time of the electron in its phase, naturally enabling electron-light synchronization in the kicker. The drift is designed such that an electron with the desired reference energy reaches the kicker at the zero crossing of the pickup wave. In that case, faster (slower) electrons will meet the wave at a decelerating (accelerating) phase, leading to cooling of the total energy spread. (a), (b), and (c) show the relative electron-light overlap after the pickup, drift, and kicker, respectively. 

To model this three-stage process, we treat both the electron and the light quantum mechanically. The electron moves in the $z$ direction with average velocity $v_0$. The optical mode has frequency $\omega$ and wavenumber in the $z$ direction $k$. We assume that the optical mode satisfies the phase-matching condition $\omega=kv_0$. We write the electron momentum operator in the scaled form $\hat{m}=\hat{p}/\hbar k$, where $\hat p$ is the momentum operator relative to the nominal momentum. We consider a single electron active in the system at a time. We estimate the validity of this assumption in the Supplementary Material (SM) \footnote{See Supplemental Material, which includes Refs.~\cite{di2019probing,garcia2021optical,stein1981estimation,talebi2020strong,karnieli2023jaynes,karnieli2024strong,sirotin2024quantum,synanidis2024quantum,leonhardt1995measuring,leonhardt1997measuring} for more detailed derivations.}. Passage through the pickup, drift, or kicker can be modeled using the following scattering operator, which evolves the entangled electron-light state (see SM for derivation):
\begin{equation}
    \hat S(g,\kappa) = e^{-i\kappa \hat{m}^2+(g\hat{b}\hat{a}^\dag-g^*\hat b^\dag \hat a)},
\end{equation}
where $\kappa=\frac{L\hbar k^2}{2\gamma_0^3m_ev_0}$ is the normalized section length and $g=-\frac{eLE_z^*}{\hbar\omega}$ the vacuum coupling strength, with $E_z$ the single mode electric field strength. $\hat b=e^{-ik\hat{z}}$ is a ladder operator which reduces the electron momentum by $\hbar k$, and $\hat a$ is the annihilation operator of the optical mode. Taking the no-recoil approximation in the pickup and kicker, the three-stage OSC transformation can be represented by the combined scattering operator
\begin{align}
    \hat S_T(g_1,g_2,\kappa)&=\hat S(g_2,0)\hat S(0,\kappa)\hat S(g_1,0)\\
    &=e^{g_2\hat b\hat a^\dag-H.c.}e^{-i\kappa\hat{m}^2}e^{g_1\hat b\hat a^\dag-H.c.}.
\end{align}
$g_1$ and $g_2$ are the complex coupling factors for the pickup and kicker, respectively, and $\kappa$ is the scaled drift length between them. Recoil effects in the pickup and kicker are treated numerically in the SM. We find that most of the recoil effects can be compensated by slightly reducing the drift length $\kappa$ for expected parameters. 

To start, we consider an electron initially in a momentum eigenstate with relative momentum $p_0=\hbar k m_0$ and the photons in the vacuum state. After the kicker, the electron will have emitted some average number of photons $\avg{n_{\text{ph}}(m_0)}$ depending on its initial momentum $m_0$. We show in the SM that this is
\begin{align}
    \avg{n_{\text{ph}}(m_0)}=&|g_1|^2+|g_2|^2\nonumber\\
    &+2|g_1||g_2|e^{-2|g_1|^2\sin^2(\kappa)}\cos(2\kappa m_0+\psi),
\end{align}
where $\psi=\Delta\phi_g-\kappa-|g_1|^2\sin(2\kappa)$, with $g_1g_2^*=|g_1||g_2|e^{i\Delta\phi_g}$. Neglecting briefly the fluctuations of the emitted photon number, the final scaled momentum of the electron is simply $m_f(m_0)=m_0-\avg{n_{\text{ph}}(m_0)}$. The lowest-order cooling condition is that the final momentum is independent of the initial momentum: $\frac{dm_f}{dm_0}|_{m_0=0}=0$, or $4\kappa|g_1||g_2|e^{-2|g_1|^2\sin^2(\kappa)}\sin(\psi)=-1$. All scaled parameters are positive, so we must choose $\sin(\psi)<0$. Furthermore, the second-order curvature in the final phase space, $\frac{d^2m_f}{dm_0^2}|_{m_0=0}$, vanishes for $\psi=-\frac{\pi}{2}$. This choice constrains the relative phase of the couplings $\Delta\phi_g$. This implies the lowest-order cooling condition
\begin{align}
    4\kappa|g_1||g_2|e^{-2|g_1|^2\sin^2(\kappa)} = 1.
\end{align}
To obtain more precise conditions for optimal cooling, we consider an electron whose initial momentum is distributed according to the distribution function $f(m)=\frac{1}{\sqrt{2\pi}\sigma_{m_0}}e^{-\frac{m^2}{2\sigma_{m_0}^2}}$, with $\sigma_{m_0}=\sigma_{E_0}/\hbar\omega$ the rms energy spread in units of the photon energy. The final energy spread in this case has two contributions: (1) electrons with different initial momentum emit a different number of photons on average, and (2) the number of emitted photons fluctuates even for definite initial momentum. 
We evaluate both contributions as a function of arbitrary parameters in the SM, including an exact expression for the final energy spread as a function of arbitrary $|g_1|$, $|g_2|$, $\psi$, and $\kappa$. For $\psi=-\pi/2$ we find an exact optimal value for $|g_2|$:
\begin{align}
    |g_2|_{\text{opt}} = \frac{4\kappa|g_1|e^{-2\kappa^2\sigma_{m_{\text{eff}}}^2}(\sigma_{m_0}^2+|g_1|^2\text{sinc}(\kappa)\cos(\kappa))}{1+2|g_1|^2\chi},
\end{align}
where $\chi=1-e^{-8\kappa^2\sigma_{m_{\text{eff}}}^2+8|g_1|^2\sin^4(\kappa)}\cos(2\kappa-8|g_1|^2\cos(\kappa)\sin^3(\kappa))$, with $\sigma_{m_{\text{eff}}}^2=\sigma_{m_0}^2+|g_1|^2\text{sinc}^2(\kappa)$ the effective energy spread after the pickup and drift. At this optimal value, the final energy spread takes the relatively simple form
\begin{align}\label{eqn:spread_at_g2opt}
    \sigma_{m_f}^2 =& \sigma_{m_0}^2\left[1-4\kappa|g_1||g_2|e^{-2\kappa^2\sigma_{m_{\text{eff}}}^2}\right]\nonumber\\
    &+|g_1|^2\left[1-4\kappa|g_1||g_2|e^{-2\kappa^2\sigma_{m_{\text{eff}}}^2}\text{sinc}(\kappa)\cos(\kappa)\right].
\end{align}
As we show in the SM, for $\kappa\ll 1$ and $|g_1|\gg 1$, this permits an analytic optimal value of the drift length
\begin{align}\label{eqn:kappa_opt}
    \kappa_{\text{opt}} \simeq \frac{1}{2\sigma_{m_{\text{eff}}}}\left(\frac{3}{1+4|g_1|^2}\right)^{1/6}.
\end{align}
At this condition the energy spread is approximately $\sigma_{m_f}^2\simeq\frac{3^{2/3}}{4\times 2^{1/3}}\frac{\sigma_{m_0}^2+|g_1|^2}{|g_1|^{4/3}}$, which is minimized for $|g_1|_{\text{opt}}=\sqrt{2}\sigma_{m_0}$ at value $\sigma_{m_f,opt}\simeq \sqrt{\frac{3^{5/3}}{8}}\sigma_{m_0}^{1/3}$. In eV, this means $\sigma_{E_f,opt}\simeq0.88(\hbar\omega)^{2/3}\sigma_{E_0}^{1/3}$. Lower photon energies are preferred for a given input energy spread, though this implies larger couplings $|g_1|$ and longer pickup and kicker sections.

\begin{figure}[h!]
    \centering
    \includegraphics[width=\linewidth]{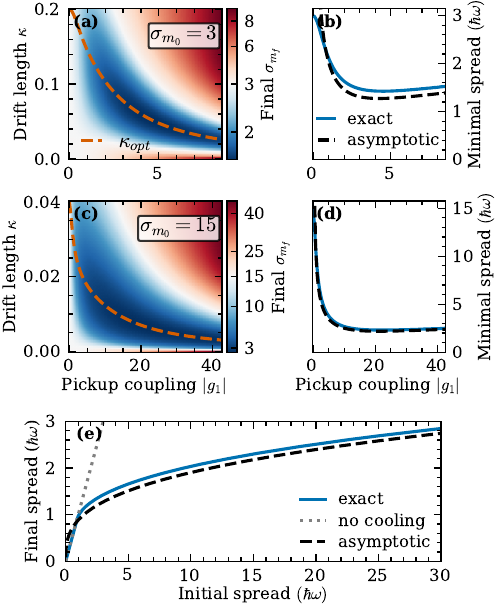}
    \caption{Expected single-pass cooling behavior using analytic derivations. (a) and (c) show final energy spread as a function of pickup coupling and drift length for $\sigma_{m_0}=3$ and $15$, respectively. (b) and (d) show the optimal final energy spread as a function of the pickup coupling compared to asymptotic estimates. (e) shows the optimal final spread as a function of the initial spread alongside the asymptotic form $\sigma_{m_f}\simeq\sqrt{\frac{3^{5/3}}{8}}\sigma_{m_0}^{1/3}$. }
    \label{fig:case_study}
\end{figure}

Figure~\ref{fig:case_study} gives an overview of the single-pass cooling performance using Eqn.~\eqref{eqn:spread_at_g2opt}. Panel~(a) shows the final spread $\sigma_{m_f}$, in units of the photon energy, as a function of the pickup coupling $|g_1|$ and drift length $\kappa$ for $\sigma_{m_0}=3$ -- an initial rms energy spread of 3 photon energies. The white region coincides with $\sigma_{m_f}=\sigma_{m_0}$, the blue represents cooling, and the red represents heating. The orange line shows the optimal drift length estimate according to Eqn.~\eqref{eqn:kappa_opt}. The estimate agrees very well with the false color plot for all pickup couplings. Panel~(b) shows the minimal energy spread as a function of the pickup coupling, both exactly from panel~(a) and using the asymptotic formula $\sigma_{m_f}^2\simeq\frac{3^{2/3}}{4\times 2^{1/3}}\frac{|g_1|^2+\sigma_{m_0}^2}{|g_1|^{4/3}}$. The agreement is quite good despite the moderate values of both $\kappa$ and $|g_1|$. Cooling down to $\sigma_{m_f}\simeq1.425$ photons, less than 50\% of the input spread, is possible. Panels~(c) and (d) show the same plots for $\sigma_{m_0}=15$. Cooling down to $\sigma_{m_f}\simeq2.3$ is possible in this case. Importantly, panels~(b) and (d) show that while the optimum coupling is $|g_1|\simeq \sqrt{2}\sigma_{m_0}$, significant cooling can still be achieved with much smaller coupling strengths. The optimal couplings are $4.24$ and $21.2$, respectively, while substantial cooling can still be accomplished with $|g_1|=2$ and $5$, respectively. Panel~(e) reports the minimal final spread as a function of the initial spread alongside the asymptotic estimate $\sigma_{m_f,opt}\simeq\sqrt{\frac{3^{5/3}}{8}}\sigma_{m_0}^{1/3}$, which agrees well. The dotted gray line shows the condition $\sigma_{m_f}=\sigma_{m_0}$ for reference. This shows that cooling is only possible if $\sigma_{m_0}\gtrsim1$, i.e., the initial energy spread exceeds the cooling photon energy. 

\textit{Numerical tracking} --- The analytic results of the last section demonstrate that significant energy spread reduction is possible, but it does not distinguish between true phase space cooling and conservative phase space manipulation. To that end, we numerically propagate the initial electron density matrix to validate the analytic results and evaluate the final phase space area. We represent the electron in the basis of momentum eigenstates with initial density matrix coefficients $\rho_{e,i}(m,m')$ and the photons in the vacuum state. We trace out the photon degrees of freedom from the final state as they will not be observed, yielding the final electron density matrix
\begin{align}
    \rho_{e,f}(m,m')=\sum_j \rho_{e,i}(m+j,m'+j)S_{m+j,j}S^*_{m'+j,j},
\end{align}
where $S_{m,n}=\langle n|\hat D(g_2)e^{-i\kappa(m-\hat{a}^\dag\hat a)^2}\hat D(g_1)|0\rangle$, with $\hat D(g)=e^{g\hat a^\dag-g^*\hat a}$ the normal bosonic displacement operator and $\ket{n}$ the $n$ excitation Fock state, see the SM for details. The final momentum spectrum is $\rho_{e,f}(m,m)$, which depends only on the input energy spectrum and not on any off-diagonal coherence information, including any chirp in the initial phase space. Determining whether true cooling has been achieved can be done by evaluating the purity of the final state $P=\text{Tr}[\rho^2]$. A pure state has $P=1$, while a highly mixed state has $P$ approaching zero. In this context, cooling means increasing the purity of the final state.

We consider an electron which starts in a statistical mixture of Gaussian states with coherent FWHM energy spread 0.1 eV, incoherent spread 0.7 eV, and incoherent FWHM duration 400 fs, similar to recent measurements from ultrafast electron microscopes \cite{fang2026quantum}. Definitions and details of the implementation are provided in the SM. We consider a case with cooling photon energy $\hbar\omega=20$ meV, corresponding to 4.84 THz light. The initial rms energy spread then corresponds to roughly 15 photons. 

\begin{figure}[h!]
    \centering
    \includegraphics[width=\linewidth]{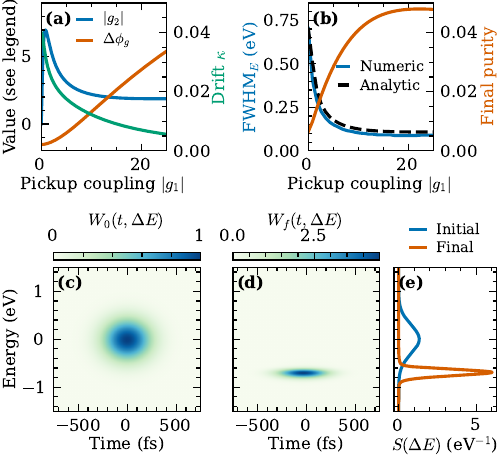}
    \caption{Numerical evaluation of cooling a $\sim0.7$ eV FWHM spread with $\hbar\omega=20$ meV photons. Panel (a) shows the optimal values of the kicker coupling, phase shift, and drift length as a function of pickup coupling. Panel (b) shows the final FWHM spread and state purity for different $|g_1|$. Panels (c) and (d) show the initial and final Wigner distributions of the electron for $|g_1|=5$, with panel (e) showing the energy spectra.}
    \label{fig:numerical}
\end{figure}

Figure~\ref{fig:numerical} summarizes the numerical results. Panel~(a) reports the optimal kicker coupling, pickup-kicker phase shift, and drift length as a function of $|g_1|$ per the analytic estimates. For strong pickup coupling, $|g_2|$ plateaus around 2, while $\kappa$ gets smaller as more photons are emitted in the pickup. In panel~(b), we evaluate the cooled energy spread and final state purity as a function of the pickup coupling using the numerical approach. Consistent with the analytic formulae, the optimal conditions are forgiving. Although the optimal $|g_1|$ is around 20, the final energy spread changes gradually for $|g_1|\geq 5$. For $|g_1|=5$, the final FWHM energy spread is already around 150 meV while the purity has grown by more than a factor of four from the initial value of $6.4\times 10^{-3}$ up to $2.9\times 10^{-2}$. This confirms that $|g_1|_{\text{opt}}\simeq\sqrt{2}\sigma_{m_0}$ is a conservative estimate, and significantly smaller couplings can be used. Panels~(c)-(d) show the Wigner distribution of the initial and final state for $|g_1|=5$. Cooling is evident from the shrinking of the overall area of the Wigner distribution from panel (c) to panel (d), reflecting the increase in the purity of the state.

\textit{Discussion} --- This single-pass approach relies on achieving coupling factors $|g|>1$ at photon energies below the electron energy spread, generally sub-0.5 eV in modern microscopes. This implies operation in the mid-to-far-IR and THz regimes. Coupling between 0.1 and 1 has been achieved at optical frequencies \cite{adiv2023observation,arend2025electrons}, and couplings above 1 are theoretically allowed but undemonstrated \cite{yang2018maximal,di2024toward,xie2025maximal,zhao2025upper}. Most work towards high coupling has focused in the optical and near-IR regimes because of overlap with available sources and interest in photon-induced near-field electron microscopy \cite{barwick2009photon,park2010photon,feist2015quantum,shiloh2022quantum}, in which the photon energy should exceed the electron energy spread. On the other hand, designing and fabricating nanophotonic structures with high coupling in the far-IR and THz regimes should be easier, as the required feature sizes are larger and the near-fields decay over longer distances. For photon transport and optics between the pickup and kicker, this scheme may benefit from significant recent advances in THz \cite{wang2004metal,anthony2011thz,leitenstorfer20232023} and CO$_2$ laser \cite{harrington2000review,temelkuran2002wavelength,ventura2020extruded} optics.

We have so far neglected optical losses, which may increase the required couplings and act as a source of decoherence. We consider lossy transport in the SM, finding corrected formulas for the optimal parameters as well as final energy spread $\sigma_{E_f}\simeq0.88\eta^{-1/3}(\hbar\omega)^{2/3}\sigma_{E_0}^{1/3}$ in the strong coupling regime, with $\eta$ the relative pickup-to-kicker power transmission. For the example in Fig.~\ref{fig:numerical}, the final energy spread remains below 250 meV as long as the corrected values are used and the transmission stays above 30\%. Lower couplings may be possible by introducing an amplifier between the pickup and kicker, which can help overcome losses and lower the required couplings. Although we focus here on single-pass cooling enabled by strong coupling, multi-pass operation with lower coupling could also be pursued in dedicated multi-pass electron microscopes \cite{kruit2016designs,juffmann2017multi,koppell2019design}, or in compact setups leveraging recent advances in on-chip electron resonators \cite{zimmermann2021charged,seidling2024resonating}. Such multi-pass geometries could even explore transverse-longitudinal coupling optics to distribute the cooling across the whole six-dimensional phase space. 

An important experimental consideration is the physical drift length required between the pickup and kicker. For the example explored above, $\kappa\simeq 0.01$ is required, corresponding to drift lengths of $1.4$ cm, $21.5$ cm, and $4.64$ m at 5, 30, and 200 keV beam energies, respectively. This suggests that a free space drift could be sufficient for cooling in SEM geometries where the beam energy rarely exceeds 30 keV. For TEMs, temporarily lower energy sections or active dispersive elements, something analogous to the chicanes used in particle accelerators, could be considered.

In conclusion, we have shown that stochastic cooling, in which entropy is removed from the electron and left behind in a discarded optical mode, can be applied to electron microscopes. High-efficiency electron-light coupling in nanophotonic structures makes it possible to achieve significant cooling in a single-pass geometry compatible with electron microscope columns, and notably the scheme is laser-free. This requires vacuum coupling strengths $|g|\gtrsim1$ in the mid-to-far IR or THz regimes along with reasonable ($\gtrsim 30\%$) transport efficiency from the pickup to the kicker in a total length of a few tens of centimeters for SEM kinetic energies. In this case, cooling of the typical energy spread of typical modern sources around 0.7 eV FWHM down to the 100 meV level is achievable. While energy spread reduction is the most direct signature of the effect, it is reflected in a commensurate increase in the purity of the quantum state of the electron and a reduction of the longitudinal emittance. While we focused on using few-THz radiation to cool down to the 100 meV level, first experiments can leverage higher photon energies to demonstrate the concept. Multi-pass schemes can not only lower the required coupling coefficients, but also open the possibility of full six-dimensional phase space cooling. Overall, our work paves a path towards enhanced beam brightness in electron microscopes, even towards fundamental quantum limits. 

\textit{Acknowledgments} --- This work was supported by the Gordon and Betty Moore Foundation (GBMF11473) and the DFG Leibniz Prize (DFG HO 4543/5-1). The authors thank Dr.\,Zhexin Zhao for helpful discussions and Hommelhoff group members for the support. 

\bibliography{apssamp}

\newpage
\onecolumngrid

\begin{center}
    {\large\textbf{Supplemental Material}}
\end{center}

\section{Derivation of the scattering operator}

The general starting point for the interaction of the electron with a photonic mode is the minimally coupled Dirac Hamiltonian:
\begin{align}
    H = \hbar \omega \hat a^\dag \hat a + c\vec\alpha\cdot\pqty{\hat {\vec p}+e\hat{ \vec{A}}} + \beta m_ec^2
\end{align}
where $\hat a$ is the optical annihilation operator with $\hbar\omega$ the photon energy, $\alpha_i=\begin{pmatrix}
    0& \sigma_i\\\sigma_i&0
\end{pmatrix}$ with $\sigma_i$ the ith Pauli matrix, $\beta=\begin{pmatrix}
    \mathbb{I}&0\\0&-\mathbb{I}
\end{pmatrix}$, $\hat{\vec{p}}$ the electron momentum operator, and $\hat{\vec{A}}$ the quantized vector potential of the optical mode. $m_e$ and $c$ are the electron mass and the speed of light, respectively. As shown in numerous other works \cite{di2019probing,garcia2021optical}, if the electron energy spread is narrow around some central velocity $v_0$ in the $z$ direction, the optical field is weak so that ponderomotive effects can be neglected, and the electron does not diffract significantly over the interaction length, this can be reduced to the effective 1D Schrödinger Hamiltonian
\begin{align}
    H = v_0 \hat p+\frac{\hat p^2}{2\gamma_0^3m_e} + \hbar\omega \hat a^\dag \hat a + ev_0\hat A_z(\hat z)
\end{align}
where $v_0=c\sqrt{1-\frac{1}{\gamma_0^2}}$ and we now use $\hat p$ to represent the z momentum operator, specifically the deviation from the reference momentum $\gamma_0m_ev_0$. The first two terms are the interaction-free energy of the electron, the third is the interaction-free energy of the optical mode, and the final term gives rise to interactions. We use the following convention for the optical vector potential:
\begin{align}
    \hat A_z(\hat z) = \frac{i}{\omega}\pqty{E_ze^{ik\hat z}\hat a-E_z^*e^{-ik\hat z}\hat a^\dag}
\end{align}
where $k$ is the longitudinal wavenumber of the mode and $E_z$ is the single-photon electric field. We move to the interaction picture with $H_0=v_0\hat p+\hbar\omega \hat a^\dag\hat a$, yielding the interaction picture Hamiltonian 
\begin{align}
    V_I = \frac{\hat p^2}{2\gamma_0^3m_e}+\frac{iev_0}{\omega}\pqty{E_ze^{ik\hat z}\hat ae^{i(kv_0-\omega)t} - E_z^*e^{-ik\hat z}\hat a^\dag e^{-i(kv_0-\omega)t}}
\end{align}
For simplicity, we assume the electron-optical interaction is phase-matched with $\omega=kv_0$. In this case, the interaction picture Hamiltonian is time-independent, and the scattering operator is exactly
\begin{align}
    \hat S(g,\kappa) = \exp[-i\kappa \hat m^2 + \pqty{ge^{-ik\hat z}\hat a^\dag - g^*e^{ik\hat z}\hat a}]
\end{align}
where we have introduced the scaled variables $\hat m=\hat p/\hbar k$, $\kappa=L\hbar k^2/2\gamma_0^3m_ev_0$, and $g=-eLE_z^*/\hbar\omega$, with $L$ the interaction length. This scattering operator describes just as well a drift with no coupling by setting $g=0$ as it does a coupling section. In most of the text we will take the no-recoil approximation in the pickup and kicker, in which $\kappa$ is set to zero in the coupling regions. This leads to the scattering operator for the three-stage OSC transformation as we write it in the text:
\begin{align}
    \hat S_T = \hat S(g_2,0)\hat S(0,\kappa)\hat S(g_1,0)
\end{align}

\section{Average photon number and variance}

The scattering operator preserves the total excitation operator $\hat m+\hat a^\dag \hat a$, reflecting energy-momentum conservation. It is most easily analyzed in the joint operator $\hat A=\hat ae^{ik\hat z}$ which simultaneously destroys a photon and increases the electron momentum by one quantum of photon momentum. Let $\ket{m_0}\otimes\ket{0}$ represent the separable state with electron momentum $\hbar k m_0$ and the optical mode in the vacuum state. For this initial state, the scattering operator can only access states $\ket{m_0-n}\otimes\ket{n}$ with $n$ photons and electron momentum $\hbar k(m_0-n)$. In this subspace, $\hat m$ behaves identically to $m_0-\hat A^\dag\hat A$. Note that $\hat A^\dag \hat A=\hat a^\dag \hat a$ simply counts the number of photons, and therefore the number of exchanged quanta, so the initial state behaves like the bosonic vacuum state. Within this subspace, we may write the scattering operator as
\begin{align}
    \hat S(g,\kappa) = \exp[-i\kappa(m_0-\hat A^\dag\hat A)^2 + (g\hat A^\dag - g^*\hat A)]
\end{align}
For $\kappa=0$, this is exactly the bosonic displacement operator $\hat D(g)=e^{g\hat A^\dag-g^*\hat A}$. So, we may write the total scattering transformation within this subspace as 
\begin{align}
    \hat S_T(m_0) = \hat D(g_2)e^{-i\kappa(m_0-\hat A^\dag\hat A)^2}\hat D(g_1)
\end{align}
We proceed in the Heisenberg picture and determine how the annihilation operator transforms under this operation. For convenience we define the transformed annihilation operator $\hat Z(m_0)=\hat S_T^\dag(m_0)\hat A\hat S_T(m_0)$. The displacement operator $\hat D(g)$ sends the annihilation operator to $\hat A+g$, so 
\begin{align}
    \hat Z (m_0)= \hat D^\dag(g_1)e^{i\kappa(m_0-\hat A^\dag\hat A)^2}(\hat A+g_2)e^{-i\kappa(m_0-\hat A^\dag \hat A)^2}\hat D(g_1)
\end{align}
Now using the normal-ordering identity $\hat Af(\hat A^\dag \hat A)=f(\hat A^\dag\hat A + 1)\hat A$, we have
\begin{align}
    \hat Z(m_0) =& \hat D^\dag(g_1)\pqty{e^{i\kappa(2m_0-1)}e^{-2i\kappa\hat A^\dag\hat A}\hat A + g_2} \hat D(g_1)\\
    =& e^{i\kappa(2m_0-1)}e^{-2i\kappa(\hat A^\dag+g_1^*)(\hat A+g_1)}(\hat A+g_1)+g_2
\end{align}
Acting on the initial state, we have 
\begin{align}
    \hat Z(m_0)\ket{0} = \pqty{e^{i\kappa(2m_0-1)}e^{-2i\kappa(\hat A^\dag+g_1^*)(\hat A+g_1)}g_1 + g_2}\ket 0 = \pqty{e^{i\kappa(2m_0-1)}\hat D^\dag(g_1)e^{-2i\kappa\hat A^\dag\hat A}\hat D(g_1)g_1 + g_2}\ket 0
\end{align}
Then for example, the expectation value of the excitation number operator $\hat{N}(m_0)=\hat Z^\dag(m_0)\hat Z(m_0)$ is 
\begin{align}
    \langle \hat N(m_0)\rangle &= |g_1|^2 + |g_2|^2 + \pqty{g_1g_2^*e^{i\kappa(2m_0-1)}\bra{g_1}  e^{-2i\kappa\hat{A}^\dag\hat{A}}\ket{g_1}+c.c.}\\
    &= |g_1|^2 + |g_2|^2 + \pqty{g_1g_2^*e^{i\kappa(2m_0-1)-i|g_1|^2\sin(2\kappa)-2|g_1|^2\sin^2(\kappa)}+c.c.}\\
    &= |g_1|^2 + |g_2|^2 + 2|g_1||g_2|e^{-2|g_1|^2\sin^2(\kappa)}\cos(2\kappa m_0+\psi)
\end{align}
where we have defined $\psi=\Delta\phi_g-\kappa-|g_1|^2\sin(2\kappa)$ with $g_1g_2^*=|g_1||g_2|e^{i\Delta\phi_g}$.

Now, for the variance we can also evaluate $\hat Z^2\ket 0$:
\begin{align}
    \hat Z^2(m_0)\ket 0 = \hat D^\dag(g_1)\pqty{g_2^2 + e^{i\kappa(2m_0-1)}e^{-2i\kappa\hat A^\dag\hat A}\pqty{ g_1^2e^{i\kappa(2m_0-3)}e^{-2i\kappa\hat A^\dag \hat A} + 2g_1g_2}}\hat D(g_1)\ket 0
\end{align}
Defining the phase $\Phi(m_0)\equiv 2\kappa m_0+\psi$, the expectation value of $\hat Z^{\dag 2}\hat Z^2$ follows as 
\begin{align}
    \langle \hat Z^{\dag 2}\hat Z^2\rangle =&  |g_2|^4 +4|g_1|^2|g_2|^2+|g_1|^4+4|g_1||g_2|e^{-2|g_1|^2\sin^2(\kappa)}\bqty{|g_2|^2\cos(\Phi) + |g_1|^2\cos(\Phi-2\kappa)}\nonumber \\
    &+ 2|g_1|^2|g_2|^2e^{-2|g_1|^2\sin^2(2\kappa)} \cos\pqty{2\Phi-2\kappa+8|g_1|^2\cos(\kappa)\sin^3(\kappa)} 
\end{align}
The photon number fluctuation is 
\begin{align}
    \sigma_{n_{\text{eff}}}^2(m_0) =& \langle\hat Z^{\dag 2}\hat Z^2\rangle+\langle\hat N\rangle - \langle\hat N\rangle ^2\\
    =& \langle \hat N(m_0)\rangle + 2|g_1|^2|g_2|^2\bqty{1 - 2e^{-4|g_1|^2\sin^2(\kappa)}\cos^2(\Phi) + e^{-2|g_1|^2\sin^2(2\kappa)}\cos\pqty{2\Phi-2\kappa+8|g_1|^2\cos(\kappa)\sin^3(\kappa)}}\nonumber\\
    &+4|g_1|^3|g_2|e^{-2|g_1|^2\sin^2(\kappa)}\bqty{\cos(\Phi-2\kappa)-\cos(\Phi)}
\end{align}
The first term is simply Poissonian noise; the final term arises from recoil in the drift and vanishes for $\kappa\to 0$.

The results so far are valid for an electron sitting initially in a momentum eigenstate. The final scaled momentum of the electron is given by $m_f=m_0-n_{\text{eff}}(m_0)$, and we would now like to understand how the initial statistical distribution of $m_0$ translates into the final distribution of $m_f$. We have 
\begin{align}
    Var(m_f) = Var(m_0) + Var(n_{\text{eff}}) - 2Cov(m_0,n_{\text{eff}}(m_0))
\end{align}
Now, the variance of $n_{\text{eff}}$ follows from the law of total variance:
\begin{align}
    Var(n_{\text{eff}}) = E[Var(n_{\text{eff}}|m_0)] + Var(E[n_{\text{eff}}|m_0])
\end{align}
In a fully classical calculation, $Var(n_{\text{eff}}|m_0)$ is zero, so this first term accounts for quantum fluctuations in the photon emission process. The second term, on the other hand, accounts for the electron emitting different average photon numbers depending on its initial energy and should recover classical results. $E[n_{\text{eff}}|m_0]$ is exactly $\langle n_{\text{eff}}(m_0)\rangle$ that we calculated before, while $Var(n_{\text{eff}}|m_0)$ is $\sigma_{n_{\text{eff}}}^2(m_0)$. The remaining expectation value and variance must be taken over the statistical distribution of $m_0$, which we assume to be Gaussian according to $f(m_0)=\frac{1}{\sqrt{2\pi}\sigma_{m_0}}e^{-\frac{m_0^2}{2\sigma_{m_0}^2}}$. This yields, for the classical part,
\begin{align}
    Var(E[n_{\text{eff}}|m_0]) = &2|g_1|^2|g_2|^2e^{-4|g_1|^2\sin^2(\kappa)}\pqty{1-e^{-4\kappa^2\sigma_{m_0}^2}}\pqty{1-e^{-4\kappa^2\sigma_{m_0}^2}\cos(2\psi)}
\end{align}
and 
\begin{align}
    E[Var(n_{\text{eff}}|m_0)] =& |g_1|^2 + |g_2|^2 + 2|g_1||g_2|e^{-2|g_1|^2\sin^2(\kappa)-2\kappa^2\sigma_{m_0}^2}\cos(\psi)\nonumber\\
    &+ 2|g_1|^2|g_2|^2\left[1 - e^{-4|g_1|^2\sin^2(\kappa)}\pqty{1+e^{-8\kappa^2\sigma_{m_0}^2}\cos(2\psi)} \right.\nonumber\\
    &\left.+ e^{-2|g_1|^2\sin^2(2\kappa)}e^{-8\kappa^2\sigma_{m_0}^2}\cos\pqty{2\psi-2\kappa+8|g_1|^2\cos(\kappa)\sin^3(\kappa)}\right]\nonumber\\
    &+4|g_1|^3|g_2|e^{-2|g_1|^2\sin^2(\kappa)}e^{-2\kappa^2\sigma_{m_0}^2}\bqty{\cos(\psi-2\kappa)-\cos(\psi)}
\end{align}
The covariance $Cov(m_0,n_{\text{eff}}(m_0))$ can be calculated from Stein's Lemma \cite{stein1981estimation}, which says that for a normally distributed random variable, $Cov(m_0,n_{\text{eff}}(m_0))=Var(m_0)E[n_{\text{eff}}'(m_0)]$, where
\begin{align}
    E[n_{\text{eff}}'(m_0)] = -4\kappa|g_1||g_2|e^{-2|g_1|^2\sin^2(\kappa)}e^{-2\kappa^2\sigma_{m_0}^2}\sin(\psi)
\end{align}
A particularly instructive way to write the final energy spread is
\begin{align}\label{eqn:exact_spread}
    \sigma_{m_f}^2 = &\sigma_{m_0}^2\pqty{1+4\kappa|g_1||g_2|e^{-2|g_1|^2\sin^2(\kappa)-2\kappa^2\sigma_{m_0}^2}\sin(\psi)}^2 \nonumber\\
    &+ 2|g_1|^2|g_2|^2e^{-4|g_1|^2\sin^2(\kappa)}\bqty{\pqty{1-e^{-4\kappa^2\sigma_{m_0}^2}}\pqty{1-e^{-4\kappa^2\sigma_{m_0}^2}\cos(2\psi)}-8\sigma_{m_0}^2\kappa^2e^{-4\kappa^2\sigma_{m_0}^2}\sin^2(\psi)}\nonumber\\
    &+ E[Var(n_{\text{eff}}|m_0)]
\end{align}
With this decomposition, the first term motivates the linear-order cooling condition $4\kappa|g_1||g_2|e^{-2|g_1|^2\sin^2(\kappa)-2\kappa^2\sigma_{m_0}^2}\sin(\psi)=-1$ as discussed in the text, with a correction $e^{-2\kappa^2\sigma_{m_0}^2}$. The second term reflects the residual phase space nonlinearity after cancellation of the first term, and the third term reflects the quantum mechanical fluctuations of the emitted photon number. This exact expression is quadratic in the kicker coupling $|g_2|$, so there is an explicit optimal value in terms of arbitrary $|g_1|$, $\kappa$, and $\psi$. The complex dependence on the remaining parameters prevents further analytical optimization without some assumptions. 

Equation~\eqref{eqn:exact_spread} is exact within the approximations leading to our scattering operator. We can use it to numerically determine optimal cooling conditions in the place of full numerical density matrix tracking. Before that, we can introduce some simplifications and approximations to suggest useful working points. Just assuming $\psi=-\pi/2$ yields the still exact but greatly simplified expression
\begin{align}
    \sigma_{m_f}^2 =& \sigma_{m_{0}}^2 + |g_1|^2- 8\kappa |g_1||g_2|e^{-2\kappa^2\sigma_{m_{\text{eff}}}^2}\bqty{\sigma_{m_{0}}^2+|g_1|^2\text{sinc}(\kappa)\cos(\kappa)}\nonumber\\
    &+|g_2|^2\bqty{1+2|g_1|^2\pqty{1-e^{-8\kappa^2\sigma_{m_{\text{eff}}}^2+8|g_1|^2\sin^4(\kappa)}\cos(2\kappa-8|g_1|^2\cos(\kappa)\sin^3(\kappa))}}
\end{align}
where $\sigma_{m_{\text{eff}}}=\sqrt{\sigma_{m_0}^2+|g_1|^2\text{sinc}^2(\kappa)}$. In this form, it becomes clear that there is an unambiguous optimal kicker coupling
\begin{align}
    |g_2|_{\text{opt}} = \frac{4\kappa |g_1|e^{-2\kappa^2\sigma_{m_{\text{eff}}}^2}\bqty{\sigma_{m_{0}}^2+|g_1|^2\text{sinc}(\kappa)\cos(\kappa)}}{1+2|g_1|^2\pqty{1-e^{-8\kappa^2\sigma_{m_{\text{eff}}}^2+8|g_1|^2\sin^4(\kappa)}\cos(2\kappa-8|g_1|^2\cos(\kappa)\sin^3(\kappa))}}
\end{align}
Evaluated at this optimal value, 
\begin{align}
    \sigma_{m_f}^2(|g_2|_{\text{opt}}) =& \sigma_{m_{0}}^2+ |g_1|^2- 4\kappa |g_1||g_2|e^{-2\kappa^2\sigma_{m_{\text{eff}}}^2}\bqty{\sigma_{m_{0}}^2+|g_1|^2\text{sinc}(\kappa)\cos(\kappa)}
\end{align}
In the high-coupling regime, $\kappa$ generally remains much less than 1, in which case we can approximate 
\begin{align}
    |g_2|_{\text{opt}} \simeq \frac{4\kappa\sigma_{m_{\text{eff}}}^2|g_1|e^{-2\kappa^2\sigma_{m_{\text{eff}}}^2}}{1+2|g_1|^2\pqty{1-e^{-8\kappa^2\sigma_{m_{\text{eff}}}^2}}}
\end{align}
and
\begin{align}
    \sigma_{m_f}^2(|g_2|_{\text{opt}})\simeq \sigma_{m_{\text{eff}}}^2\bqty{1-\frac{16\kappa^2\sigma_{m_{\text{eff}}}^2|g_1|^2}{\text{cosh}(4\kappa^2\sigma_{m_{\text{eff}}}^2)+(1+4|g_1|^2)\text{sinh}(4\kappa^2\sigma_{m_{\text{eff}}}^2)}}
\end{align}
Defining $x=4\kappa^2\sigma_{m_{\text{eff}}}^2$ and $G=1+4|g_1|^2$, this is minimized as a function of $x$ when 
\begin{align}
    \frac{xG-1}{G-x}=\tanh(x)
\end{align}
Now, for $\kappa\ll1$ we can estimate $\tanh(x)\simeq x - \frac{x^3}{3}$, while
\begin{align}
    \frac{Gx-1}{G-x}\simeq -\frac{1}{G} + \pqty{1-\frac{1}{G^2}}x+\pqty{\frac{1}{G}-\frac{1}{G^3}}x^2\simeq -\frac{1}{G}+x
\end{align}
where we've assumed both that $x\ll 1$ and $G\gg 1$. The condition then reduces to $x^*=(3/G)^{1/3}$. This predicts the optimal drift
\begin{align}
    \kappa_{\text{opt}}\simeq \frac{1}{2\sigma_{m_{\text{eff}}}}\pqty{\frac{3}{1+4|g_1|^2}}^{1/6}
\end{align}
Plugging this into the energy spread and taking $|g_1|\gg 1$ leads to the lowest-order expression,
\begin{align}
    \sigma_{m_f}^2(|g_2|_{\text{opt}},\kappa_{\text{opt}})\simeq\frac{3^{2/3}}{4\times 2^{1/3}}\frac{\sigma_{m_0}^2+|g_1|^2}{|g_1|^{4/3}}
\end{align}
This is minimized as a function of $|g_1|$ for $|g_1|_{\text{opt}}=\sqrt{2}\sigma_{m_0}$, for which it takes the value $\sigma_{m_f,opt}=\sqrt{\frac{3^{5/3}}{8}}\sigma_{m_0}^{1/3}$.

\section{Description of mixed electron states}

We consider the initial electronic state to be a statistical mixture of generally chirped Gaussian wavepackets. A chirped Gaussian wavepacket with central momentum $\hbar k m_0$, chirp $C$, and scaled longitudinal position $\phi_0=kz_0$ can be described by the superposition of momentum eigenstates 
\begin{align}
    \ket{m_0,\phi_0} = \int dm \left[\frac{1}{2\pi\sigma_{m_c}^2}\right]^{1/4}\exp\left[-\frac{(m-m_0)^2}{4\sigma_{m_c}^2}(1+iC)-i(m-m_0)\phi_0\right]\ket{m}\equiv \int dm\psi_p(m;m_0,\phi_0)\ket{m}
\end{align}
where $\sigma_{m_c}$ is the coherent rms energy spread. Furthermore, we can describe the statistical fluctuations of the wavepackets using the distribution function
\begin{align}
    f_s(m_0,\phi_0)=\frac{1}{2\pi\epsilon_{m_i}}\exp\left[-\frac{\sigma_{\phi_i}^2m_0^2-2\sigma_{m_i\phi_i}m_0\phi_0+\sigma_{m_i}^2\phi_0^2}{2\epsilon_{m_i}^2}\right]
\end{align}
where $\sigma_{m_i}$ and $\sigma_{\phi_i}$ are the incoherent energy and phase spreads, respectively, $\sigma_{m_i\phi_i}$ representing the incoherent chirp, and $\epsilon_{m_i}^2=\sigma_{m_i}^2\sigma_{\phi_i}^2-\sigma_{m_i\phi_i}^2$ the longitudinal emittance. The density matrix of the resulting statistical mixture of chirped wavepackets is 
\begin{align}
    \rho_e =& \int dm_0 d\phi_0 f_s(m_0,\phi_0)\ket{m_0,\phi_0}\bra{m_0,\phi_0}\\
    =&\int dm_0d\phi_0 f_s(m_0,\phi_0)\int dm dm' \psi_p(m;m_0,\phi_0)\psi_p^*(m';m_0,\phi_0)\ket{m}\bra{m'}\\
    =& \int dm dm' \rho_e(m,m')\ket{m}\bra{m'}
\end{align}
where the density matrix elements are 
\begin{align}
    \rho_e(m,m') =& \frac{1}{\sqrt{2\pi(\sigma_{m_c}^2+\sigma_{m_i}^2)}}\exp\left[-\frac{(m+m')^2+2i(m^2-m'^2)(C+2\sigma_{m_i\phi_i})}{8(\sigma_{m_c}^2+\sigma_{m_i}^2)}\right]\nonumber\\
    &\times\exp\left[-\frac{(m-m')^2\left((1+C^2)\sigma_{m_i}^2+4\sigma_{m_c}^4\sigma_{\phi_i}^2+\sigma_{m_c}^2\left(1+4\epsilon_{m_i}^2-4C\sigma_{m_i\phi_i}\right)\right)}{8\sigma_{m_c}^2(\sigma_{m_c}^2+\sigma_{m_i}^2)}\right]
\end{align}
Though we include the chirp parameter here, it does not impact the spectrum $\rho(m,m)$ and therefore plays no role in the single-pass cooling of the energy spread.

\section{Multi-electron effects and valid current range}

Our analysis is valid when a single electron is active in the cooling system at a time. Multi-electron effects become important when the pickup radiation emitted by one electron can interact with another in the kicker. In that case, the radiation from one electron acts as an incoherent kick on the second. Avoiding this requires that a single electron's radiation pulse duration is shorter than the inter-electron spacing. In a periodic nanostructure, the radiated pulse length is approximately the number of periods multiplied by the radiation wavelength. Taking for example Smith-Purcell-like radiators, the phase-matching condition for radiation emitted normal to the electron propagation direction requires $\beta_0=\Lambda/\lambda$, with $\lambda$ the radiation wavelength and $\Lambda$ the structure period. The radiation pulse duration is then $(L/\Lambda)\lambda/c=L/v_0$. The average separation between electrons in a beam with current $I$ is $e/I$. The single-active-electron condition is then $IL \ll ev_0$. This implies an upper limit of $IL\simeq 1.58$ nA$\cdot$cm at 30 keV kinetic energy, and $IL\simeq 3.34$ nA$\cdot$cm at 200 keV. Assuming structure lengths less than 1 cm, currents in the nA range should pose no issue, and higher currents can be possible depending on the exact design. Cooling is still possible with multi-electron effects included, just at a reduced rate. A detailed analysis of these effects is beyond our current scope. 

\section{Recoil effects in pickup and kicker and their compensation}

In the analysis thus far we have ignored the effects of recoil in the pickup and kicker. With such large couplings, it could be important to take this into account. We can do so by using the full scattering matrix:
\begin{align}
    \hat S_T &= \hat S(g_2,\kappa_2)\hat S(0,\kappa)\hat S(g_1,\kappa_1)\\
    &= e^{-i\kappa_2\hat m^2 + (g_2\hat b\hat a^\dag - g_2^*\hat b^\dag \hat a)} e^{-i\kappa\hat m^2} e^{-i\kappa_1\hat m^2 + (g_1\hat b\hat a^\dag - g_1^*\hat b^\dag \hat a)}
\end{align}
The inclusion of recoil in the pickup and kicker has two effects. The lowest-order effect is simply to lengthen the effective drift from $\kappa$ to something like $\kappa+\kappa_1+\kappa_2$. This is in fact an overestimate, since the electron accumulates energy changes over the length of the structure. A reasonable estimate for the effective increased drift length might be $\kappa_{\text{eff}}=\kappa+(\kappa_1+\kappa_2)/2$. Beyond an effective drift length correction, recoil can also qualitatively change the behavior of the electron-light interaction in the pickup and kicker \cite{talebi2020strong,karnieli2023jaynes,karnieli2024strong,sirotin2024quantum,synanidis2024quantum}. The first effect can be compensated simply by shifting the $\kappa$ used in the drift by the correction term $(\kappa_1+\kappa_2)/2$.

\begin{figure}[h!]
    \centering
    \includegraphics[width=\linewidth]{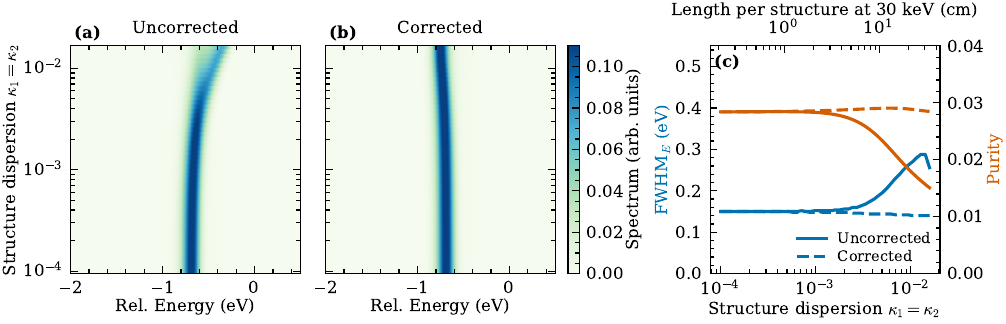}
    \caption{Recoil effects in the pickup and kicker, and their compensation with effective drift correction for $\kappa_1=\kappa_2$. Panels (a) and (b) show the final spectrum as a function of the scaled structure length without and with drift length compensation. Panel (c) shows the final FWHM energy spread and state purity as a function of the same.}
    \label{fig:recoil_compensation}
\end{figure}

Figure~\ref{fig:recoil_compensation} showcases the effects of including recoil in the numerical calculations and their compensation with a corrected drift length. The parameters are the same as in the main text with $|g_1|=5$, only now we consider identical structure lengths for simplicity with $\kappa_1=\kappa_2$, which can be converted to physical lengths when a kinetic energy is specified. Panels (a) and (b) show the final energy spectrum as a function of the structure length. In (a), the drift length is uncorrected and for longer structure lengths, the spectrum shifts further from the nominal position and becomes distorted. In panel (b), we reduce the drift length by $(\kappa_1+\kappa_2)/2$ to compensate for the extra dispersion in the pickup and kicker. We see that the spectrum is much less affected by the pickup and kicker recoil in this case. Panel (c) shows the final FWHM energy spread and state purity as a function of the structure lengths. We also plot the true structure length at 30 keV kinetic energy on the top x axis for reference. In the uncorrected case, the final spread grows and the purity shrinks for $L>2$ cm. When the drift length is corrected, this degradation is significantly weaker, and the tolerable structure length can be extended to $\simeq 40$ cm. Beyond this, the corrected drift length drops below zero, becoming unphysical. $40$ cm structure length is already impractical anyways, so we conclude that for any practical design, recoil effects are easily correctable with small adjustments to the drift length. 

\begin{figure}[h!]
    \centering
    \includegraphics[width=0.75\linewidth]{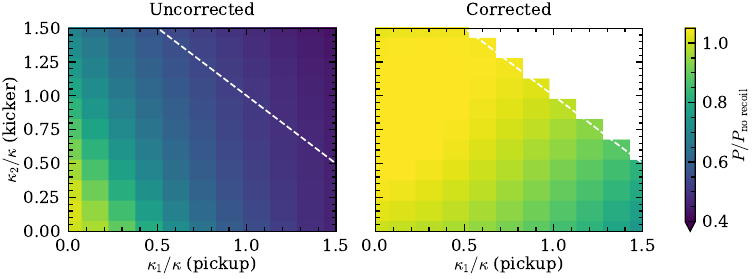}
    \caption{Final purity as a function of scaled lengths of the pickup ($\kappa_1$) and kicker ($\kappa_2$) with and without recoil correction of the drift length.}
    \label{fig:recoil_map}
\end{figure}

For further reference, in Figure~\ref{fig:recoil_map} we independently scan $\kappa_1$ and $\kappa_2$ and plot the resulting purity of the state with and without recoil correction. The white dashed line marks the region beyond which the corrected drift becomes negative. Again we see that the purity is minimally affected when correction is applied for a large range of pickup and kicker scaled lengths. The performance is more sensitive to recoil in the pickup than the kicker, but across the entire region the purity is maintained above at least 75\% of the no-recoil value. We also note that our correction $(\kappa_1+\kappa_2)/2$ is a simple estimate, and even better performance could be possible with careful optimization. 

\section{Impact of photon loss during transport}

Loss of photons during transport from the pickup to the kicker can be modeled generally with the beam-splitter transformation (see, for example, \cite{leonhardt1995measuring,leonhardt1997measuring})
\begin{align}
    \hat a\rightarrow \sqrt{\eta}\hat a + \sqrt{1-\eta}\hat F
\end{align}
where $\hat F$ is the annihilation operator for a fictitious bosonic loss channel starting in the vacuum state, with unitarity requiring $[\hat F,\hat F^\dag]=1$. This is represented by the unitary transformation
\begin{align}
    \hat S_L = e^{\theta(\hat a^\dag \hat F - \hat a\hat F^\dag)}
\end{align}
where $\cos(\theta)=\sqrt{\eta}$. The total transformation is then $\hat S_T=\hat S(g_2,0)\hat S_L\hat S(0,\kappa)\hat S(g_1,0)$, and since $\hat S_L$ commutes with the drift transformation we can think of it as being applied either after the pickup or before the kicker. If we take the latter, we can note that
\begin{align}
    \hat S(g_2,0)\hat S_L = \hat S_L\hat S_L^\dag \hat S(g_2,0)\hat S_L = \hat S_L\exp\left[g_2\hat b\left(\sqrt{\eta}\hat a^\dag+\sqrt{1-\eta}\hat F^\dag\right)-H.c.\right]
\end{align}
Now, $\hat b$, $\hat a$, and $\hat F$ all commute with each other, and further $\hat b$ commutes with $\hat b^\dag$, so this can be factorized as 
\begin{align}
    \hat S(g_2,0)\hat S_L = \hat S_L\hat S_F(\sqrt{1-\eta}g_2,0)\hat S(\sqrt{\eta}g_2,0)
\end{align}
where we've written $\hat S_F(g,0)$ as equivalent to $\hat S(g,0)$, only acting on the $\hat F$ subspace instead of $\hat a$. 

In this form, the total transformation before application of $\hat S_F$ is exactly the same as the original one, only now with $g_2$ scaled to an effective value $\sqrt{\eta}g_2$. The whole transformation is then followed by another unitary transformation which mixes the electron-photon Hilbert space with the Hilbert space of the lossy mode, leading to decoherence. We may write the total lossy scattering operator as 
\begin{align}
    \hat S_{T,L}=\hat S_L\hat S_F(\sqrt{1-\eta}g_2,0)\hat S_T(g_1,\sqrt{\eta}g_2,\kappa)
\end{align}
The density matrix at the end of the loss-corrected OSC transformation is an entangled state between the electrons and the primary photonic mode, separably multiplied by the vacuum state for the loss mode $\ket{0}\bra{0}_F$. The final electronic state is the trace over both $\hat a$ and $\hat F$ degrees of freedom:
\begin{align}
    \rho_{e,f} = \text{Tr}_{a,F}[\hat S_L\hat S_F(\rho_{e,a}\otimes\ket{0}\bra{0}_F)\hat S_F^\dag\hat S_L^\dag] = \text{Tr}_{a,F}[\hat S_F(\rho_{e,a}\otimes\ket{0}\bra{0}_F)\hat S_F^\dag]
\end{align}
$\hat{S}_F$ is completely independent of $\hat a$, so the trace over $\hat a$ passes directly onto the density matrix, yielding the photon-traced-out electronic density matrix of the normal OSC transformation with scaled $|g_2|$. Writing this as $\rho_{e,f}=\int dm dm'\rho_{e,f}(m,m')\ket m\bra{m'}$, and letting $\hat S_F\ket{m}\otimes\ket{0}_F=\sum_l D_{l,0}\ket{m-l}\otimes \ket{l}_F$, we find the final electronic density matrix after tracing out the loss subsystem
\begin{align}
    \rho_{e,f} = \int dmdm' \left[\sum_l \rho(m+l,m'+l)\left|D_{l,0}\left(g_2\sqrt{1-\eta}\right)\right|^2\right]\ket m\bra {m'}
\end{align}
Conveniently, the matrix elements for the displacement operator starting from vacuum have a simple closed form:
\begin{align}
    |D_{l,0}(\alpha)|^2 = e^{-|\alpha|^2}\frac{|\alpha|^{2l}}{l!}
\end{align}
so we have 
\begin{align}
    \rho_{e,f} = \int dmdm' \left[\sum_l \rho(m+l,m'+l)e^{-|g_2|^2(1-\eta)}\frac{(|g_2|^2(1-\eta))^{l}}{l!}\right]\ket m\bra {m'}
\end{align}
The final result is a discrete convolution of the final OSC density matrix with a Poissonian of mean $(1-\eta)|g_2|^2$. Such a convolution simply shifts the mean of the spectrum by $(1-\eta)|g_2|^2$ and adds the same to the final variance. It follows that the final energy spread is, for $\psi=-\pi/2$, 
\begin{align}
    \sigma_{m_f}^2 =& \sigma_{m_{0}}^2 + |g_1|^2- 8\kappa |g_1|\sqrt{\eta}|g_2|e^{-2\kappa^2\sigma_{m_{\text{eff}}}^2}\bqty{\sigma_{m_{0}}^2+|g_1|^2\text{sinc}(\kappa)\cos(\kappa)}\nonumber\\
    &+\eta|g_2|^2\bqty{\frac{1}{\eta}+2|g_1|^2\pqty{1-e^{-8\kappa^2\sigma_{m_{\text{eff}}}^2+8|g_1|^2\sin^4(\kappa)}\cos(2\kappa-8|g_1|^2\cos(\kappa)\sin^3(\kappa))}}
\end{align}
the optimal $|g_2|$ then follows as 
\begin{align}
    |g_2|_{\text{opt}} = \frac{1}{\sqrt{\eta}}\frac{4\kappa |g_1|e^{-2\kappa^2\sigma_{m_{\text{eff}}}^2}\bqty{\sigma_{m_{0}}^2+|g_1|^2\text{sinc}(\kappa)\cos(\kappa)}}{\frac{1}{\eta}+2|g_1|^2\pqty{1-e^{-8\kappa^2\sigma_{m_{\text{eff}}}^2+8|g_1|^2\sin^4(\kappa)}\cos(2\kappa-8|g_1|^2\cos(\kappa)\sin^3(\kappa))}}
\end{align}
and as before 
\begin{align}
    \sigma_{m_f}^2 = \sigma_{m_{0}}^2 + |g_1|^2- 4\kappa |g_1|\sqrt{\eta}|g_2|_{\text{opt}}e^{-2\kappa^2\sigma_{m_{\text{eff}}}^2}\bqty{\sigma_{m_{0}}^2+|g_1|^2\text{sinc}(\kappa)\cos(\kappa)}
\end{align}
As before, for $\kappa\ll 1$,
\begin{align}
    \sqrt{\eta}|g_2|_{\text{opt}} \simeq \frac{4\kappa\sigma_{m_{\text{eff}}}^2|g_1|e^{-2\kappa^2\sigma_{m_{\text{eff}}}^2}}{\frac{1}{\eta}+2|g_1|^2\pqty{1-e^{-8\kappa^2\sigma_{m_{\text{eff}}}^2}}}
\end{align}
and
\begin{align}
    \sigma_{m_f}^2(|g_2|_{\text{opt}})\simeq \sigma_{m_{\text{eff}}}^2\bqty{1-\frac{16\kappa^2\sigma_{m_{\text{eff}}}^2\eta|g_1|^2}{\text{cosh}(4\kappa^2\sigma_{m_{\text{eff}}}^2)+(1+4\eta|g_1|^2)\text{sinh}(4\kappa^2\sigma_{m_{\text{eff}}}^2)}}
\end{align}
At this point, this is identical to before with the replacement $|g_1|^2\rightarrow\eta|g_1|^2$, though not in $\sigma_{m_{\text{eff}}}^2$, which makes sense, as that noise is induced before any loss. Note also that this immediately gives a precise limit on transmission to achieve cooling, given by the point where $\sigma_{m_f}^2=\sigma_{m_0}^2$:
\begin{align}
    \eta_{min} = \frac{e^{4\kappa^2\sigma_{m_{\text{eff}}}^2}}{16\kappa^2\sigma_{m_{\text{eff}}}^4-4|g_1|^2\sinh(4\kappa^2\sigma_{m_{\text{eff}}}^2)}
\end{align}

The same optimum follows from before as long as $1+4\eta|g_1|^2$ remains reasonably large:
\begin{align}
    \kappa_{\text{opt}}\simeq\frac{1}{2\sigma_{m_{\text{eff}}}}\left(\frac{3}{1+4\eta|g_1|^2}\right)^{1/6}
\end{align}
Now the asymptotic energy spread follows as 
\begin{align}
    \sigma_{m_f}^2\simeq \frac{3^{2/3}}{4\times 2^{1/3}}\frac{\sigma_{m_0}^2+|g_1|^2}{\eta^{2/3}|g_1|^{4/3}}
\end{align}
The optimal value of the pickup coupling is unaffected, it is still $|g_1|=\sqrt{2}\sigma_{m_0}$, only now the optimal energy spread is $\sigma_{m_f}=\sqrt{\frac{3^{5/3}}{8}}\eta^{-1/3}\sigma_{m_0}^{1/3}$.

\begin{figure}[h!]
    \centering
    \includegraphics[width=\linewidth]{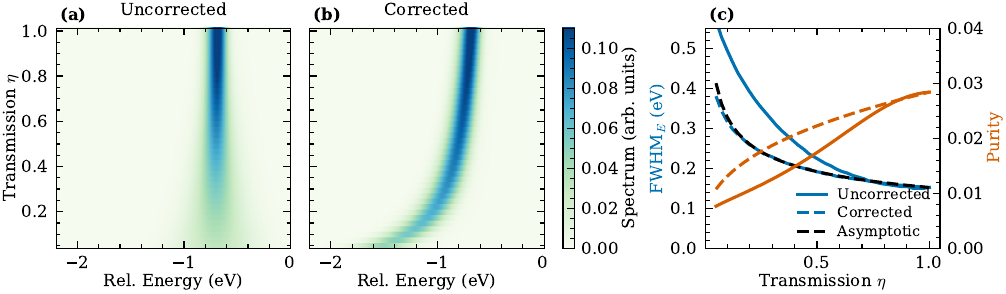}
    \caption{Study of loss effects on cooling performance. Panels (a) and (b) show the uncorrected and corrected spectra as a function of the pickup-to-kicker transmission $\eta$. Panel (c) summarizes the final energy spread and state purity as a function of the transmission. }
    \label{fig:loss_effects}
\end{figure}

Figure~\ref{fig:loss_effects} explores the effect of photon loss between the pickup and kicker on the same case study from the main text. Panel (a) shows the spectrum as a function of transmission, reflecting both the reduction in effective kicker coupling to $\sqrt{\eta}g_2$ and the added noise from decoherence into the loss mode, and with no direct correction for the loss. Panel (b) shows the spectrum now using the corrected optimal values for $|g_2|$ and $\kappa$. It is clear that the spectrum remains narrower down to lower pickup-kicker transmissions when using the corrected optimal values, and the net energy decreases further as the optimal $g_2$ grows like $1/\sqrt{\eta}$. Panel (c) shows the final energy spread and state purity as a function of the transmission for the uncorrected (solid) and corrected (dashed) cases. For the energy spread, we also plot the asymptotic energy spread estimate scaling as $\eta^{-1/3}$. Using loss-corrected optimal values for the drift length and kicker coupling maintains good cooling performance even down to around 30\% transmission.

\end{document}